\documentclass{aa}  
\usepackage{graphicx}
\usepackage{txfonts}
\usepackage{bm}
\usepackage[breaklinks=true]{hyperref}
\usepackage{nccmath}   
\usepackage{amsmath}

\begin{document}

\title{Three-dimensional MHD wave propagation near a coronal null point: a new wave mode decomposition approach}
\titlerunning{3D MHD wave propagation near null}

  \author{
  N. Yadav
          \inst{ \footnotemark}
          \and
          Rony Keppens\inst{}  
          \and 
          B. Popescu Braileanu\inst{}
          }
   \institute{
  Centre for mathematical Plasma Astrophysics, Department of Mathematics, KU Leuven, \\Celestijnenlaan 200B, 3001 Leuven, Belgium\\
              \email{nitnyadv@gmail.com}
             }
   \date{}
 
  \abstract
   {
Ubiquitous vortex flows at the solar surface excite magnetohydrodynamic (MHD) waves that propagate to higher layers of the solar atmosphere. In the solar corona, these waves frequently encounter magnetic null points.
The interaction of MHD waves with a coronal magnetic null in realistic 3D setups requires an appropriate wave identification method.
   }
   {
   We present a new MHD wave decomposition method that overcomes the limitations of existing wave identification methods.
   Our method allows to investigate the energy fluxes in different MHD modes at different locations of the solar atmosphere as waves generated by vortex flows travel through the solar atmosphere and pass near the magnetic null.}
   {We use the open-source {\tt MPI-AMRVAC} code to simulate wave dynamics through a coronal null configuration. 
   We apply a rotational wave driver at our bottom photospheric boundary to mimic vortex flows at the solar surface.
   To identify the wave energy fluxes associated with different MHD wave modes, we employ a wave-decomposition method that is able to uniquely distinguish different MHD modes.
   Our proposed method utilizes the geometry of an individual magnetic field-line in 3D space to separate out velocity perturbations associated with the three fundamental MHD waves. We compare our method with an existing wave decomposition method that uses magnetic flux surfaces instead.
Over selected flux surfaces, we calculate and analyze temporally averaged wave energy fluxes, as well as acoustic and magnetic energy fluxes. 
Our wave decomposition method allows us to estimate the relative strengths of individual MHD wave energy fluxes.
   }
   {Our method for wave identification is consistent with previous flux-surface-based methods and gives expected results in terms of wave energy fluxes at various locations of the null configuration. 
   We show that ubiquitous vortex flows excite MHD waves that contribute significantly to the Poynting flux in the solar corona.
   Alfv\'en wave energy flux accumulates on the fan surface and fast wave energy flux accumulates near the null point. 
   There is a strong current density buildup at the spine and fan surface.
   }
   {The proposed method has advantages over previously utilized wave decomposition methods, since it may be employed in realistic simulations or magnetic extrapolations, as well as in real solar observations, whenever the 3D fieldline shape is known.
   The essential characteristics of MHD wave propagation near a null, such as wave energy flux accumulation and current buildup at specific locations, translate to the more realistic setup presented here.
   The enhancement in energy flux associated with magneto-acoustic waves near nulls may have important implications in the formation of jets and impulsive plasma flows.}

   \maketitle
\section{Introduction}
Turbulent convection is crucial for a variety of solar processes, including the formation and decay of magnetic concentrations, magnetic flux cancellation, and so on (\citealt{ake2009,stein2012}).
Vortices are a major component of any turbulent flow, and the Sun's highly turbulent intergranular plasma is no exception. 
Though vortex flows were first seen on the solar surface, their detection is not restricted to the  near-surface levels:
they have also been observed in the solar chromosphere (\citealt{2009A&A...507L...9W}).
Solar observations and realistic 3D magnetoconvection simulations both revealed vortices existing on a wide range of spatial and temporal scales (\citealt{kitishvili2012,yadav2020,2021A&A...645A...3Y}).
One distinguishes small-scale vortex flows (\citealt{Bonet2008,bonet2010,ParkS2016}), large-scale vortex flows (\citealt{1988brandt,Attie2009,iker2018}) or solar tornadoes (\citealt{Wedemeyer-Bohm2012}), depending on their locations and spatial extents. 
Ever smaller vortices are being detected with improved observational and simulation capabilities. 
Turbulence is thought to create vortices of varying sizes, although the exact mechanism is unknown and remains an unresolved question in solar physics (\citealt{2021A&A...649A.121B}).

In the solar atmosphere, magnetic fields serve to guide and influence vorticity propagation.
Vorticity generated by torsional motions at the surface propagates to the upper layers of the solar atmosphere in the form of torsional Alfv\'en waves (\citealt{2019NatCo..10.3504L}).
These upwardly propagating torsional MHD waves carry a lot of energy that could be dissipated in the solar atmosphere's upper layers, making vortex flows a possibly important source of solar atmospheric heating.
Various mode transformations and mode couplings occur as these torsional Alfv\'en waves propagate through the solar atmosphere.
Despite the fact that they are difficult to detect in observations due to their incompressible nature, some recent observations have confirmed their presence (\citealt{2009Sci...323.1582J,2014Sci...346D.315D,2015NatCo...6.7813M, 2017NatSR...743147S}).
Since the Alfv\'en waves are highly incompressible, in order to heat the plasma, they require a mechanism to convert their magnetic energy into thermal energy by mode conversion, phase mixing, or a non-linear turbulent cascade.
Furthermore, because Alfv\'en waves have magnetic tension as restoring force, they are less affected by large thermodynamic gradients in the transition zone than the other MHD wave modes.
Unlike acoustic waves, they have no propagation cut-off (\citealt{2007ApJ...659..650M}).

Magnetic null points, like vortices, are a common occurrence in the solar atmosphere, particularly in the solar corona.
Magnetic nulls are locations where the magnetic field strength and hence the Alfv\'en speed is effectively zero.
Because they are truly local, small-scale features, they are still impossible to observe with existing instrumentation.
Numerical calculations and extrapolations of photospheric magnetograms, on the other hand, suggest that null points will be found throughout the solar chromosphere and corona (\citealt{2009SoPh..254...51L}).
They can be classified as positive or negative nulls, based on the sign of eigenvalues in the magnetic field's Jacobian matrix.
For example, a null point is called positive if one eigenvalue is negative and the other two are positive (or have positive real parts).
The real and imaginary parts of the eigenvectors corresponding to these two  eigenvalues span a plane such that all field-lines are originating from the null point. 
These field-lines form a surface called the fan surface of the null (\citealt{1996PhPl....3..759P, 2005LRSP....2....7L}). 
Multiple null points can also develop in certain magnetic configurations: as a result of a local bifurcation a null point may split into two separate nulls, or two nulls may coalesce to form a new null (\citealt{1999RSPSA.455.3931B}).

MHD waves have been intensively investigated due to their potential as a diagnostic tool for coronal seismology and their contribution to solar atmospheric heating (e.g. see reviews by \citealt{2012RSPTA.370.3217P,2015RSPTA.37340261A,2020ARA&A..58..441N,2021JGRA..12629097S}).
Despite the existence of various wave-based heating theories, reconnection-based heating mechanisms have often been preferred since the wave energy flux is usually found to be insufficient to overcome radiative losses (\citealt{2005Natur.435..919F}).
Magnetic reconnection and heating due to waves are not independent mechanisms however, since magnetic reconnection can be a source of waves \citep{2004Verwichte, 2008Jess, 2008Luna, 2012Li, Prov2018}, and waves can produce instabilities needed to trigger reconnection processes \citep{Isobe2006, 2014Lee, 2009McLaughlin}.
Moreover, recent advances in observational instrumentation and simulation capabilities imply that wave-based heating could be a viable process for maintaining the high temperatures of the solar atmosphere (\citealt{2020A&A...642A..52A, 2021A&A...652A..43Y, 2021A&A...648A..28A}).
These new results have drawn renewed interest in wave heating mechanisms, both observationally and theoretically. Combinations of MHD wave aspects, vortices and null point ingredients have only been studied to some extent (\citealt{2011SSRv..158..205M,2013A&A...558A.127T,2013ApJ...776L...4S,2015ApJ...799....6M, 2017ApJ...837...94T, Tarr_2019,2019MNRAS.484.1390M, 2019NatCo..10.3504L,2020A&A...643A.166T,2021A&A...649A.121B}).
Because vortex flows, resulting torsional Alfv\'en waves and localized null points are so common, their co-occurrence is quite likely, opening up new challenges in the study of wave and energy propagation.

\begin{figure}
     \centering
        \includegraphics[scale=0.35,trim=0.5cm 0 3cm 0]{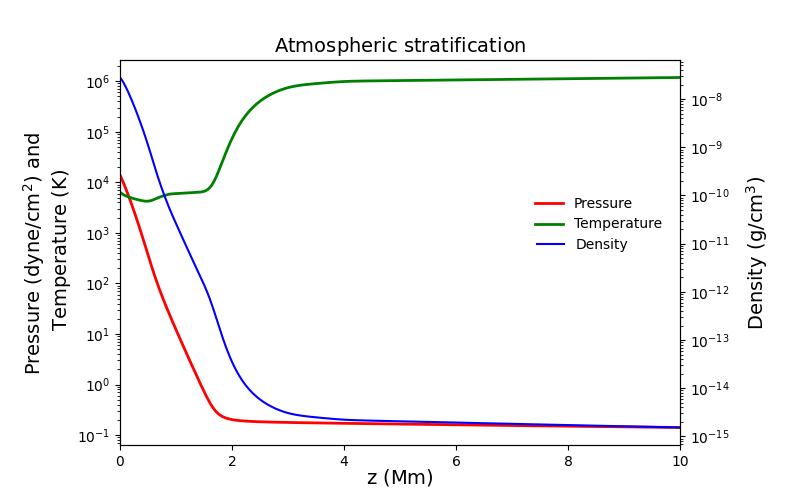}
          \caption{Height variation of temperature (green), density (blue), and pressure (red). The left vertical axis displays the temperature and the pressure, and the right vertical axis displays the mass density in CGS units.}
             \label{atmosphere}
\end{figure}
MHD wave propagation in the solar atmosphere is strongly influenced by the magnetic field topology and undergoes numerous wave transformations as it encounters obstacles such as the Alfv\'en-acoustic equipartition layer, the transition region, and magnetic nulls, among others.
There have been several modelling efforts to investigate wave conversion and wave coupling as MHD waves travel through the different layers of the solar atmosphere with/without a magnetic null (\citealt{2003ApJ...599..626B,2004A&A...420.1129M,2006ApJ...653..739K,2011A&A...531A..63G,2011AnGeo..29.1029F,2012ApJ...755...18V,2013A&A...560A..97P,2015A&A...577A..70S,2017ApJ...837...94T,Tarr_2019,2020NatSR..1015603S}).
In particular, \citet{2020NatSR..1015603S} investigated the nonlinear effects associated with interaction of an Alfv\'en wave with a magnetic null point in a 2.5D simulation setup.
They found that an incoming Alfv\'en pulse excites magneto-acoustic waves which in return, induce current density and are responsible for creation of plasma flows that are possible sources of jets in the solar corona.

Many previous studies are limited to 2D, where Alfv\'en waves are omitted as they have a perturbation normal to the plane of the adopted background magnetic field.
Inclusion of a third dimension such as in 2.5D simulations or in full 3D simulations allows Alfv\'en waves to exist.
In a 2.5D simulation, it is relatively easy to isolate the three fundamental MHD waves, as slow and fast modes can be separated by projecting velocities into the directions parallel and perpendicular to the local magnetic field in the plane of the magnetic field (at least in a low-beta plasma), while the Alfv\'en wave is along the normal to the plane.
However, in full 3D setups it becomes really challenging to isolate the three modes if wave identification is desired.
At the same time, full 3D MHD simulations have proven to be a very powerful tool to investigate 3D wave propagation and transformations in the solar atmosphere. 
As a result, numerous wave decomposition methods are in use that allow the contributions of the three basic MHD waves to be separated, which is valid under certain assumptions discussed in detail in Sec. \ref{wmd}.
Being able to do such a wave mode decomposition has important applications.
In direct solar observations, it helps us understand the role of specific wave types in various spectacularly dynamic and intricate phenomena. Depending on the type of the wave, we know that certain dissipation mechanisms (not included in our ideal MHD study here) might be more efficient. For example, in a 2D geometry, partial ionization effects are a more efficient mechanism of dissipation for fast modes than for slow modes \citep{2021A&A...653A.131P,ambi2}.
Moreover, by comparing observed wave properties to analytical/numerical models, coronal seismology helps us to determine the physical properties of the coronal regions through which the waves are propagating (\citealt{2015SSRv..190..103J}). Even nonlinearly, the ideal MHD description features various shock types that in a complicated way generalize the linear wave modes: MHD shocks form a three-parameter family of solutions, and linear slow, Alfv\'en and fast waves can be considered as shocks of infinitesimal strength, although the Alfv\'en waves more properly relate to rotational (Alfv\'en) discontinuities \citep{goedbloed_keppens_poedts_2019}.
Therefore, although in any stratified medium, there are no pure slow, Alfv\'en and fast modes, it still remains meaningful to identify how a signal decomposes in its pure MHD wave types locally.
This is done in practice by a local, mutually orthogonal decomposition, allowing to discuss wave transformations and their possible role in plasma heating.

In the present paper, we investigate the propagation of linear MHD waves generated by wave-like torsional motion at the solar surface to the higher layers of the solar atmosphere, including photosphere, chromosphere, transition region and the solar corona. 
Our magnetic field configuration resembles a local parasitic polarity magnetic field embedded within a homogeneous magnetic field of the opposite polarity. This introduces a 3D null in the coronal regions.
The goal of this research is to learn how MHD waves driven by ubiquitous vortices at the solar surface propagate across the stratified solar atmosphere, specifically in the vicinity of an isolated 3D magnetic null.
We propose a new approach for identifying the three wave modes, and compare it to previously used methods. We discuss the advantages of our method over existing strategies.
We compute the wave energy fluxes and investigate the impact of the presence of a null point on energy flux accumulation on various locations for the three MHD wave types.
We also examine the specific locations where strong currents will be generated as these will be the sites for preferential heating when resistivity is included.

The paper is organized in the following way: we describe the MHD equations and how they are solved numerically, the adopted stratification of the solar atmosphere and the specific magnetic topology in Section \ref{equations}. We discuss the proposed wave decomposition method and its comparison with existing methods in Section \ref{wmd}, and in Section \ref{results} we discuss the results related to wave energy fluxes and current accumulation. We finally conclude in Section \ref{conclusion}.
\begin{figure}
     \centering
        \includegraphics[scale=0.3,trim=1cm 0 1cm 0]{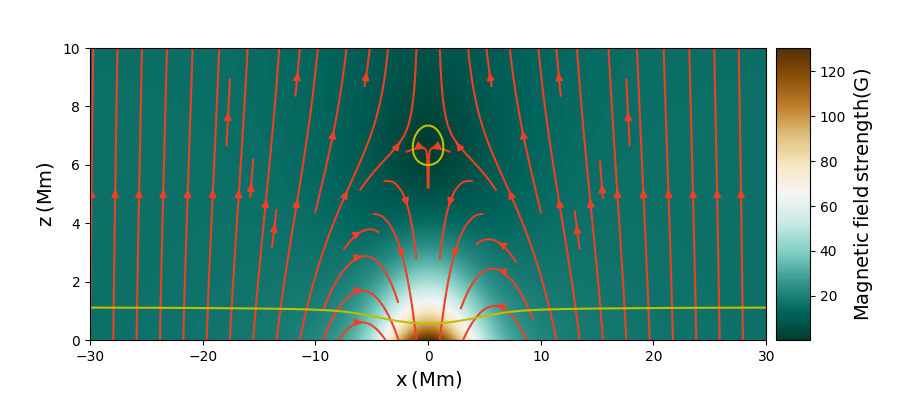}
             \caption{Spatial map of the magnetic field strength at the $y=0$ plane. The red lines represent magnetic fieldlines, and the yellow lines are contours for $\beta$=1: this happens in the chromosphere and near the null point.}
                 \label{mag_vertical_cut}
\end{figure}

\section{Numerical setup}\label{equations}
\subsection{Governing equations}
We perform three-dimensional numerical simulations of wave propagation through the stratified solar atmosphere.
The simulations are carried out with the MPI-parallelised Adaptive Mesh Refinement Versatile Advection Code ({\tt MPI-AMRVAC}; \citealt{Porth_2014,2018ApJS..234...30X,KEPPENS2021316}).
We use a single fluid MHD module of the code where a background stratified equilibrium is explicitly removed from the equations, to better simulate the linear to weakly nonlinear wave regime that is of interest here.
The well-known governing ideal MHD equations are the following:
\begin{figure}
     \centering
        \includegraphics[scale=0.45]{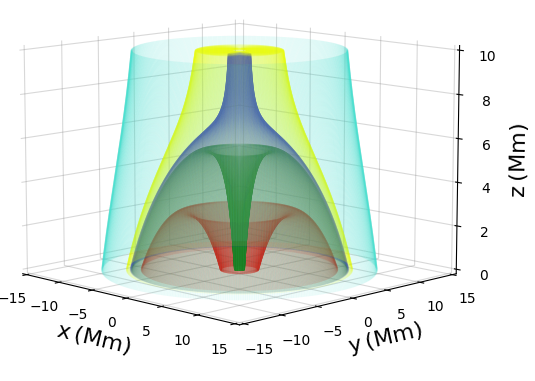}
         \caption{Three-dimensional visualization of various magnetic flux surfaces.}
            \label{magnetic_surfaces}
\end{figure}
\begin{figure}
     \centering
        \includegraphics[scale=0.375,trim=2cm 0 5cm 0]{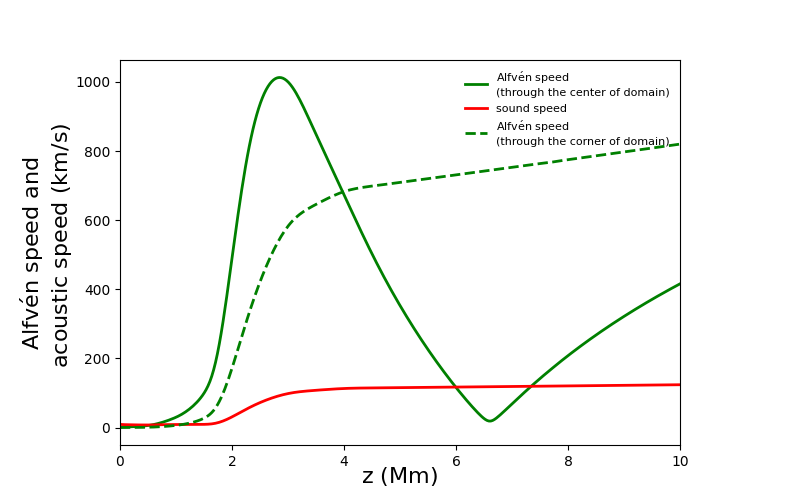}
         \caption{Height variation of Alfv\'en and sound speeds in the equilibrium atmosphere. By construction, the sound speed is only height-dependent. On the other hand, the Alfv\'en speed profile changes in a two-dimensional fashion as it depends on the magnetic field strength: we show both the central and corner variation.}
            \label{Alfven_speed}
\end{figure}

\begin{gather}
\frac{\partial \rho}{\partial t} + \nabla \cdot \left(\rho\mathbf{v}\right) = 0 \,,\\
\frac{\partial (\rho\mathbf{v})}{\partial t} + \nabla \cdot \left[\rho\mathbf{v} \mathbf{v} +  \left(p+\frac{B^2}{2}\right) \mathbb{I}  - \mathbf{B}\mathbf{B} \right]= \rho\mathbf{g} \,,\\
\frac{\partial }{\partial t} \left(e_{int}+\frac{\rho v^2}{2}+\frac{B^2}{2}\right) +  \nabla \cdot \left[\mathbf{v} \left(e_{int}+\frac{\rho v^2}{2}+B^2 +p \right) - \mathbf{B}\mathbf{B}\cdot \mathbf{v} \right ] \nonumber \\
\hspace*{2cm}= \rho\mathbf{v}\cdot \mathbf{g} \,,\\
\frac{\partial \mathbf{B}}{\partial t} +\nabla \cdot (\mathbf{v}\mathbf{B}-\mathbf{B}\mathbf{v}) = \mathbf{0} \,,
\end{gather}
where $\rho$, $p$, $\mathbf{v}$, and $\mathbf{B}$ represent the mass density, thermal pressure, fluid velocity, and magnetic field vector, respectively.
The internal energy density is
\begin{equation}
e_{int}=p/(\gamma-1)\,,
\end{equation}
where $\mathrm{\gamma=5/3}$ is the adiabatic index.
We actually split the MHD variables $\mathbf{B}$, $\rho$ and $p$ into split-off, prescribed, time-independent parts and their deviations, such that
$\rho=\rho_0 + \rho_1$,
$p=p_0 + p_1$, $\mathbf{B}=\mathbf{B}_0 + \mathbf{B}_1$, where we subtract the magneto-hydrostatic (time-independent background) equilibrium (where possibly a non-vashing current density distribution $\mathbf{J}_0=\nabla\times \mathbf{B}_0$ enters)
such that:
\begin{equation}
\frac{\partial{\rho_0}}{\partial t}=0\,,
\frac{\partial{p_0}}{\partial t}=0\,,
\frac{\partial{\mathbf{B}_0}}{\partial t}=\mathbf{0}\,,
-\nabla p_0 + \rho_0 \mathbf{g}\ + \mathbf{J}_0 \times \mathbf{B}_0=\mathbf{0},
\nonumber
\end{equation}
Then Eqs. (1\,-\,4) become: 
\begin{gather}
\frac{\partial \rho_1}{\partial t} + \nabla \cdot \left(\rho\mathbf{v}\right) = 0 \,,\\
\frac{\partial (\rho\mathbf{v})}{\partial t} + \nabla \cdot \Biggl[ \rho\mathbf{v} \mathbf{v} +  \left(p_1+\frac{B_1^2}{2}+\mathbf{B}_0\cdot\mathbf{B}_1\right) \mathbb{I}  - \nonumber \\
\left ( \mathbf{B}_0\mathbf{B}_1 + \mathbf{B}_1\mathbf{B}_0 + \mathbf{B}_1\mathbf{B}_1 \right) \Biggr] = \rho_1 \mathbf{g} \,,\\
\frac{\partial }{\partial t} \left(e_{int1}+\frac{\rho v^2}{2}+\frac{B_1^2}{2}\right) +  \nabla \cdot \Biggl[\mathbf{v} \left(e_{int}+\frac{\rho v^2}{2} +p_1 \right) \nonumber \\
-( \mathbf{v} \times \mathbf{B} ) \times \mathbf{B}_1 \Biggr ] = \rho_1\mathbf{v}\cdot \mathbf{g} -p_0 \nabla \cdot \mathbf{v}-(\nabla \times \mathbf{B}_0)\cdot(\mathbf{v}\times\mathbf{B}_1) \,,\\
\frac{\partial \mathbf{B}_1}{\partial t} +\nabla \cdot (\mathbf{v}\mathbf{B}-\mathbf{B}\mathbf{v})= \mathbf{0} \,.
\end{gather}
 \subsection{Initial condition}
The computational domain is a 3D box in Cartesian geometry $(x, y, z)$ such that $z$ corresponds to the vertical direction i.e. normal to the solar surface.
Our simulation domain extends from $-30 \leq x \leq 30 \: \mathrm{Mm}, -30 \leq y \leq 30 \: \mathrm{Mm} \: \mathrm{and} \: 0 \leq z \leq 10 \: \mathrm{Mm}$ and is resolved by $200 \times 200 \times 300$ grid points.

We use the temperature profile from VAL-C model of the solar atmosphere in the photosphere and the chromosphere (\citealt{1981ApJS...45..635V}), and join it smoothly to a constant temperature of 1 MK for the solar corona. This slightly smoothens the transition region variation (which is not represented well anyway as non-adiabatic effects like thermal conduction are ignored here), as seen in Fig.~\ref{atmosphere}. The background state is a full magnetohydrostatic equilibrium i.e. 
\begin{equation}
    -\nabla p_0 + \mathbf{J}_0 \times \mathbf{B}_0 + \rho_0 \mathbf{g} =0
\end{equation}
where $p_0$ is the background kinetic pressure, $\mathbf{g}$ is the solar gravitational acceleration and $\rho_0$ is the background density. In fact, for simplicity, we use here a potential magnetic field $\mathbf{B}_0$ so that current term $\mathbf{J}_0$ and the background Lorentz force vanishes.
The resulting pressure and density are then just one-dimensional profiles, and the height variation of pressure, density along with our input temperature profile are shown in Fig. \ref{atmosphere}.
The initial potential configuration of the atmospheric magnetic field is similar to that of \citet{Candelaresi_2016}.
It includes an isolated 3D magnetic null point located at the apex of a fan surface.
This magnetic field topology represents a parasitic polarity embedded in a dominant opposite polarity magnetic field.
The analytic expressions for the three components of magnetic field vector in dimensionless form are: 
\begin{gather}
B_x=-B_0\frac{3x(z-d)}{[x^2+y^2+(z-d)^2]^{5/2}} \,,\\
B_y=-B_0\frac{3y(z-d)}{[x^2+y^2+(z-d)^2]^{5/2}} \,,\\
B_z=B_0 + B_0\frac{[x^2+y^2-2(z-d)^2]}{[x^2+y^2+(z-d)^2]^{5/2}} \,,
\end{gather}
where $B_0$ and $d$ are the parameters that decide the magnetic field topology and the location of the magnetic null point. We used $B_0 \simeq 11.2$ G and $d=-6$ Mm
for the present simulations. 
The map of the background magnetic field strength (in real units) across a vertical cut passing through the magnetic null is displayed in Fig. \ref{mag_vertical_cut}.
To get a three-dimensional view of our magnetic topology, we numerically construct a few constant magnetic flux tubes shown by their field-line boundaries i.e. we show `flux surfaces' in Fig. \ref{magnetic_surfaces}. They enclose the same amount of magnetic flux over their whole length.
The equilibrium magnetic field is rotationally symmetric about the vertical ($z$ in our case) axis.
The propagation of MHD waves and their energy fluxes will be studied in further detail using the three inner flux surfaces depicted in red, green, and blue colors.

Height variations of Alfv\'en speed (green curves) and sound speed (red curve) are displayed in Fig. \ref{Alfven_speed}. Since the sound speed depends only on the local plasma temperature (from pressure and mass density), it is independent of the horizontal location and is only height-dependent.
In contrast, the Alfv\'en speed includes the local magnetic field that has 3D spatial structuring and hence depends both on height and horizontal position.
For illustration, we show the height variation of Alfv\'en speed through the central (solid green curve) and corner (dashed green curve) position of the domain.

\subsection{Boundary conditions and wave driver}
To avoid reflection of waves from the upper boundary we employ a uniform absorbing layer of a thickness of 1 Mm at the top boundary such that the time-dependent variables are damped exponentially in this layer:
\begin{gather}
    u^\prime =u \, \mathrm{exp}\left(-a \frac{z-z_{db}}{z_{tb}-z_{db}}\right) \,,
\end{gather}
where $z_{db}$ and $z_{tb}$ are the bottom and the top of the absorbing layer respectively, and $a$ is the damping coefficient.
This kind of absorbing layer has been successfully used in previous numerical studies related to wave propagation in the solar atmosphere (\citealt{2010ApJ...719..357F,2021A&A...653A.131P}).
Lateral boundaries are kept periodic for all variables.
The bottom boundary includes the wave driver in the horizontal velocity components, all other time dependent variables viz. $p_1$, $\rho_1$, $\mathbf{B}_1$, and $v_z$ are kept zero. 

The wave driver represents an oscillating spiral vortex at a circular region of a radius 1 Mm and has the following analytical form
\begin{gather}
   v_x=-v_0 \, \mathrm{sin}(\omega t) \, \mathrm{sin}(\phi + \phi_0) \, \sqrt{x^2+y^2} \,,\\
   v_y=v_0 \, \mathrm{sin}(\omega t) \, \mathrm{cos}(\phi + \phi_0) \, \sqrt{x^2+y^2} \,,
\end{gather}
where $v_0$=1 km/s is the amplitude, $\omega$ is the driver frequency, $\phi = \mathrm{tan}^{-1} \left(\frac{y}{x}\right)$ is the azimuth angle in the bottom plane, and $\phi_0=\frac{\pi}{10}$ is the expansion parameter of the spiral vortex flow.
This velocity driver at the bottom boundary perturbs the equilibrium to excite MHD waves. 
\begin{figure}
     \centering
        \includegraphics[scale=0.42,trim=0cm 0 0cm 0]{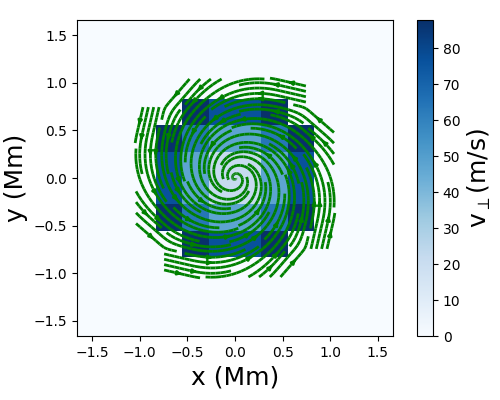}
         \caption{Spatial profile of the velocity driver applied in the ghost cells of the simulation domain at peak amplitude. Green lines are streamlines of the velocity vector field, over-plotted on the map of velocity magnitude.}
            \label{driver}
\end{figure}

We used a velocity driver with a frequency $\omega \simeq 26.2$ mHz, which equates to a time period of 240 seconds.
Since the driver frequency is much greater than the Br\"unt–V\"ais\"al\"a frequency ($\sim$ 2.2 mHz) in our corona, dispersion due to gravity will not be strong enough to change the nature of pure MHD waves.
Fig.~\ref{driver} displays the spatial profile of the velocity driver in the bottom plane at the peak amplitude. 
We used a velocity driver with a small amplitude such that despite propagating through a gravitationally stratified atmosphere, perturbations remain small compared to the background values.
We do this to focus on the linear to weakly nonlinear regime in the present study and to avoid nonlinear effects, such as wave steepening and nonlinear coupling.

Note that our driver does not use a prechosen (vertical) spatial wavelength directly, but fixes the temporal variation instead. In general, only when the wavelength is much smaller than the scale on which background quantities vary, can one approximate the background plasma to be locally uniform and MHD waves to be identifiable in their pure linear slow, Alfv\'en and fast form.
In our case, because of gravitational stratification, the wavelength of propagating waves may increase and become comparable to the local density scale height of the solar corona. 
However, our simulation domain extends to a distance of about one fifth of the scale height that is approximately 50 Mm for coronal plasma at 1 MK. Therefore, background plasma parameters remain more or less uniform in our coronal part, as can also be seen in Fig. \ref{atmosphere}.

\section{Wave mode decomposition}\label{wmd}

\subsection{MHD wave modes}
In a compressible, homogeneous and uniformly magnetized plasma of infinite extent, MHD theory supports three fundamental linear MHD waves viz. slow magneto-acoustic wave, fast magneto-acoustic wave and Alfv\'en wave. 
In an inhomogeneous setting, one can still identify the corresponding slow, Alfv\'en and fast subspectrum, which organize the entire collection of MHD eigenmodes (\citealt{goedbloed_keppens_poedts_2019}).
The three wave modes become coupled in the presence of a gravitational stratification and a non-uniform magnetic field.
In particular, magneto-acoustic waves travelling through the solar atmosphere get coupled and undergo linear mixing in the layer where the sound and the Alfv\'en speed are comparable (\citealt{2006MNRAS.372..551S,2006RSPTA.364..333C}). 
These waves, on the other hand, are mostly decoupled in places where either magnetic or thermal pressure dominate.
In such situations, one of the waves is acoustic, while the other is magnetic.
Decomposing the plasma motions into locally slow magneto-acoustic, fast magneto-acoustic, and Alfv\'en waves is essential to examine the numerous wave transformations occurring in the solar environment and to quantify wave energy fluxes that relate to the three fundamental MHD waves.

Much intuition about MHD wave properties relates to their known behavior in uniform media.
Magneto-acoustic waves have a parallel component of plasma motion and are driven by total pressure (gas pressure and magnetic pressure) and magnetic tension. 
They do not propagate parallel vorticity and have compressible plasma motions.
Alfv\'en waves are driven purely by magnetic tension and exhibit transverse plasma movements.
They propagate vorticity and have incompressible plasma motions.
Invoking the compression due to parallel and transverse velocity motions and the parallel vorticity, \citet{2017MNRAS.466..413C} constructed proxies that can be used to separate out the various wave types.
However, in the gravitationally stratified solar atmosphere and in a non-uniform plasma, these waves are no longer pure modes and can propagate both compression and vorticity (\citealt{2003ASSL..294.....G,roberts_2019}).

\begin{figure*}
     \centering
        \includegraphics[scale=0.25,trim=0 0 0 0]{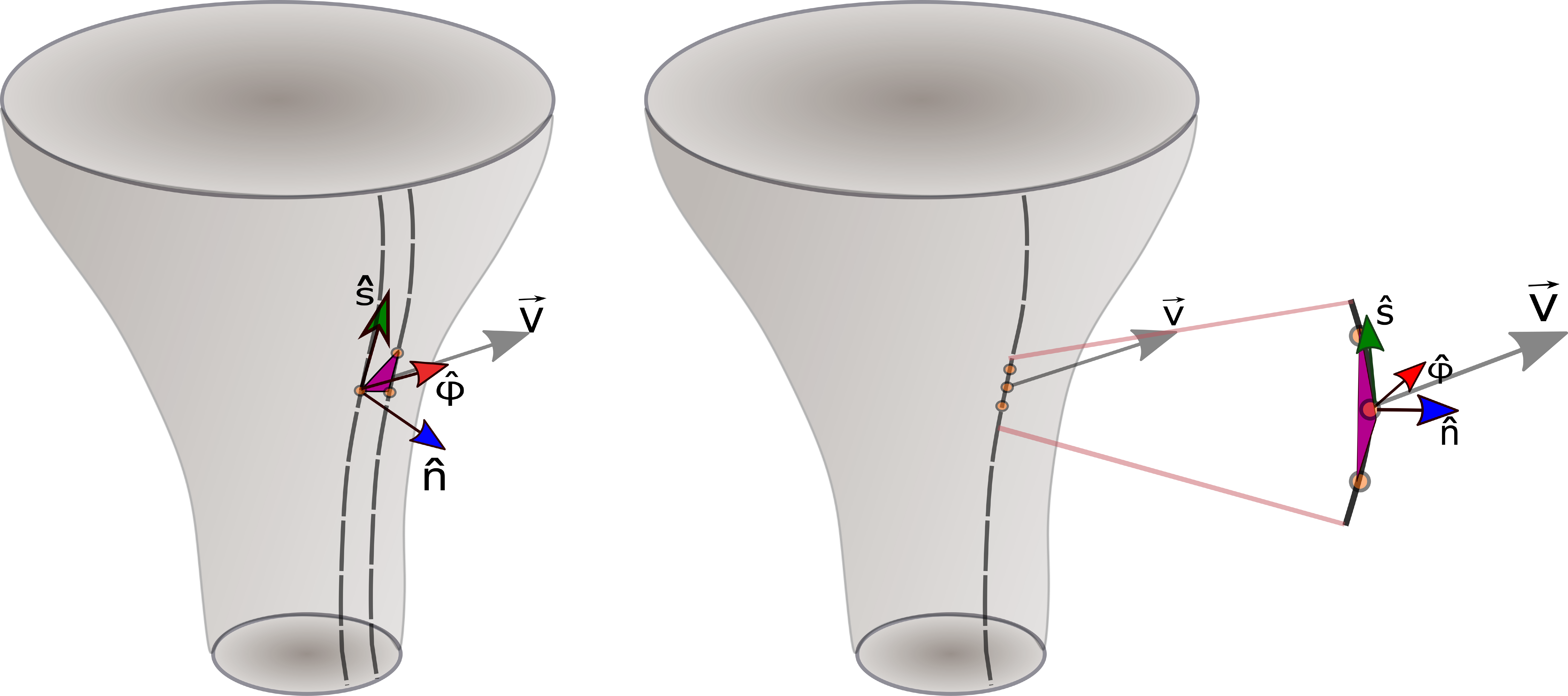}
         \caption{Cartoon representation of wave decomposition methods. Left: wave decomposition approach utilized by \citet{2015ApJ...799....6M}. Right: our proposed wave decomposition method based on the field-line geometry. The unit direction vectors for motions associated with Alfv\'en, fast magneto-acoustic, and slow magneto-acoustic waves are represented by red, blue, and green vectors, respectively, while the fluid velocity vector is represented by a grey vector.}
            \label{cartoon2}
\end{figure*}

In much of the solar chromosphere and corona, the regions we are dealing with in the present paper, magnetic pressure dominates over thermal pressure and the plasma beta is much less than unity.
In such low-beta plasma, where the Alfv\'en speed is much larger than the sound speed, magneto-acoustic oscillations can meaningfully be split into slow and fast components, where slow magneto-acoustic waves are predominantly guided along the magnetic field vector. From Fig.~\ref{Alfven_speed}, one notices that this is true over most of our domain, except in the lowest atmospheric regions and near the null.

\subsection{An overview of MHD mode decompositions}
A few approaches for isolating individual, pure MHD wave modes in low-beta plasma regions have been proposed and used in the literature, although most of them are only suitable to 2D or 2.5D systems.
In 2D simulations, it is relatively easy to separate out the wave modes, as Alfv\'en waves do not exist while slow and fast mode can be separated by taking projections of the velocity vector along and normal to the background magnetic field (\citealt{2002ApJ...564..508R,2003ApJ...599..626B, 2009ApJ...694..411K}).
In 2.5D simulations, there is an invariant direction such that the magnetic pressure gradient vanishes and the only driving force in the invariant direction is magnetic tension. 
Therefore, plasma motions along this invariant direction are associated with Alfv\'en waves. 
Fast and slow waves can be distinguished by taking projections along and normal to the background field in the same way as in 2D simulations (\citealt{2012A&A...545A...9T,2013A&A...558A.127T}).
If the magnetic field is in the $x-y$ plane, the velocity perturbations corresponding to the three MHD waves can be calculated by the projections of the velocity vector $\mathbf{v}$ as shown below: 
\begin{ceqn}
\begin{align}
    \mathrm{v}_{alf} &= \mathbf{v} \cdot \hat{\mathbf{z}} \,,\\
    \mathrm{v}_{slow} &= \mathbf{v} \cdot \hat{\mathbf{s}} \,,\\
    \mathrm{v}_{fast} &=  \mathbf{v} \cdot (\hat{\mathbf{z}} \times \hat{\mathbf{s}}) \,,
\end{align}
\end{ceqn}
where $\mathbf{v}$ is a linear velocity field on top of a magnetohydrostatic equilibrium and $\hat{\mathbf{s}}$ is the unit vector along the (in-plane) magnetic field vector.

In full 3D simulations, where the magnetic field is not confined to a 2D geometry, isolation of MHD waves becomes non-trivial as there are an infinite number of vectors normal to the magnetic field vector.
However, under certain assumptions, it is possible to derive analytical expressions for characteristic directions for velocity fluctuations associated with the different MHD modes.
A few methods widely used in the literature are discussed below.

One possible way to separate the three different MHD waves is by taking velocity projections onto the following three characteristic directions (\citealt{2011JPhCS.271a2042K,2012ApJ...746...68K}):
\begin{ceqn}
\begin{align}
    \hat{\mathbf{s}} &= [\mathrm{cos} \, \phi \: \mathrm{sin} \, \theta,\:  \mathrm{sin} \, \phi \: \mathrm{sin} \, \theta,\:  \mathrm{cos}\, \theta] \,,\\
    \hat{\bm{\phi}} &= [-\mathrm{cos} \, \phi  \:  \mathrm{sin^2}  \,\theta \:  \mathrm{sin}  \, \phi, \:  1- \mathrm{sin^2}  \, \theta \:  \mathrm{sin^2}  \, \phi, \:  -\mathrm{cos}  \, \theta \:  \mathrm{sin}  \, \theta \: \mathrm{sin} \, \phi] \,,\\
    \hat{\mathbf{n}} &=  [-\mathrm{cos}\, \theta, \: 0, \: \mathrm{cos} \, \phi \, \mathrm{sin}\, \theta] \,,
\end{align}
\end{ceqn}
where a plane parallel atmosphere is permeated by magnetic field such that $\theta$ is the inclination angle, measured from the vertical and $\phi$ is the azimuth angle, measured from the $x-z$ plane.
Here, the first projection ($\hat{\mathbf{s}}$) corresponds to the slow magneto-acoustic wave, the projection along $\hat{\bm{\phi}}$ corresponds to the Alfv\'en wave, according to the asymptotic polarization direction derived by \citet{2008SoPh..251..251C}, and the projection along $\hat{\mathbf{n}}$ selects the fast magneto-acoustic wave.
This specific decomposition method is derived assuming that the wave propagation vector is confined to the $x-z$ plane such that the $y$-component in $\hat{\mathbf{n}}$ is always zero. However, as an approximation, this decomposition has been also applied to fully three dimensional setups where the wave propagation vector is not constrained to lie in the 2D plane (\citealt{2010ApJ...719..357F}). 

\citet{2016ApJ...819L..11S} instead decomposed the different MHD wave modes in a low-beta solar atmosphere by projecting the velocity vector in the three characteristic directions defined by the following unit vectors:
\begin{ceqn}
\begin{align}
    \hat{\mathbf{s}} &= [\mathrm{cos} \, \phi \: \mathrm{sin} \, \theta, \:  \mathrm{sin} \, \phi \: \mathrm{sin} \, \theta, \: \mathrm{cos}\, \theta] \,,\\
    \hat{\bm{\phi}} &= [\mathrm{sin}\, \phi, -\mathrm{cos}\, \phi, 0] \,,\\
    \hat{\mathbf{n}} &=  [\mathrm{cos}\, \phi \: \mathrm{cos}\, \theta, \:  \mathrm{sin}\, \phi \: \mathrm{cos}\, \theta, \:  -\mathrm{sin}\, \theta] \,.
\end{align}
\end{ceqn}
This decomposition can only be used when the flux tube is symmetric w.r.t. the $z$ axis that defines the direction of gravity throughout the domain. Indeed, we see how here the Alfv\'en velocity fluctuations are strictly in the $x-y$ plane.
However, in general cases when flux tubes are inclined or twisted, this method loses merit as velocity perturbations corresponding to Alfv\'en waves can also have components along the $z$ direction.

\citet{2019A&A...625A.144R} employed a method similar to the one used by \citet{2016ApJ...819L..11S} to decompose velocity components.
However, they calculated the direction vectors numerically instead of using analytic expressions. 
For example, to calculate the direction vector for Alfv\'en waves, they calculate the iso-contour of the magnetic field for all horizontal layers and then calculate a linear fit at each pixel.
This method works perfectly when the flux tube profile is known beforehand and is symmetric about the $z$ axis. 
However, for curved magnetic flux tubes or in general for any complex 3D time-dependent magnetic field, the azimuthal direction vector calculated using isocontours of magnetic field at a fixed geometric height would not be fully consistent.

Instead, \citet{2015ApJ...799....6M} invoked the concept of ``flux tubes'' and ``flux surfaces'' to split the velocity perturbation vector into three waves.
A flux tube is a well defined theoretical construct, mostly used in analytical models where we have e.g. cylindrical (or translational) geometry, that contains a constant amount of magnetic flux at each cross-section. A flux surface is defined as any smooth surface with a local normal $\hat{\mathbf{n}}$ such that locally, $\textbf{B} \cdot \hat{\mathbf{n}} = 0$, i.e. a surface build up by individual magnetic field-lines.
To distinguish between the three MHD modes in such geometry, they took projections of the local velocity vector along three directions viz. along the magnetic field vector (for the slow magneto-acoustic wave); normal to the flux tube (for the fast magneto-acoustic wave) and along the azimuthal vector that is perpendicular to the magnetic field and parallel to the flux surface (Alfv\'en wave). 
To construct the magnetic flux surface numerically in their simulation domain they selected seed points on the circumference of a circular area and traced magnetic field-lines in 3D space.
To calculate the normal vector for their axially symmetric flux surface, they made use of neighboring field-lines and computed normal vectors by constructing polygons. 
Once the normal vector $\hat{\mathbf{n}}$ is known, the azimuthal unit vector can readily be calculated as $\hat{\bm{\phi}}=\hat{\mathbf{s}} \times \hat{\mathbf{n}}$.
This procedure is shown in Fig.~\ref{cartoon2} in a cartoon representation. 
This triplet of mutually orthogonal directions has also been used for detailed MHD spectroscopy of inhomogeneous 2.5D configurations, e.g., fusion devices (\citealt{goedbloed_keppens_poedts_2019}) and 2.5D flux ropes in stratified atmospheres (\citealt{2011A&A...532A..93B,2011A&A...532A..94B}).

The direction vectors calculated by the aforementioned method depend on the chosen flux surface, therefore, we would refer to this method as `flux-surface method'.
\citet{2015ApJ...799....6M} constructed axially symmetric magnetic flux surfaces, appropriate due to their axisymmetric magnetic field configuration.
However, in reality, a magnetic configuration can be significantly more complex e.g., in fully 3D magneto-convection simulations as well as in actual solar observations.
Observations of solar active regions, for example, show that the overall emerging flux is the result of several small flux bundles emerging concurrently or in succession, rather than a single cohesive, monolithic flux bundle (\citealt{1998ApJ...492..804E}).
Therefore, the choice of flux surface can greatly influence these direction vectors and hence, the velocity components corresponding to MHD waves.
\begin{figure*}
     \centering
        \includegraphics[scale=0.2,trim=0 -0.5cm 0 -0.5cm]{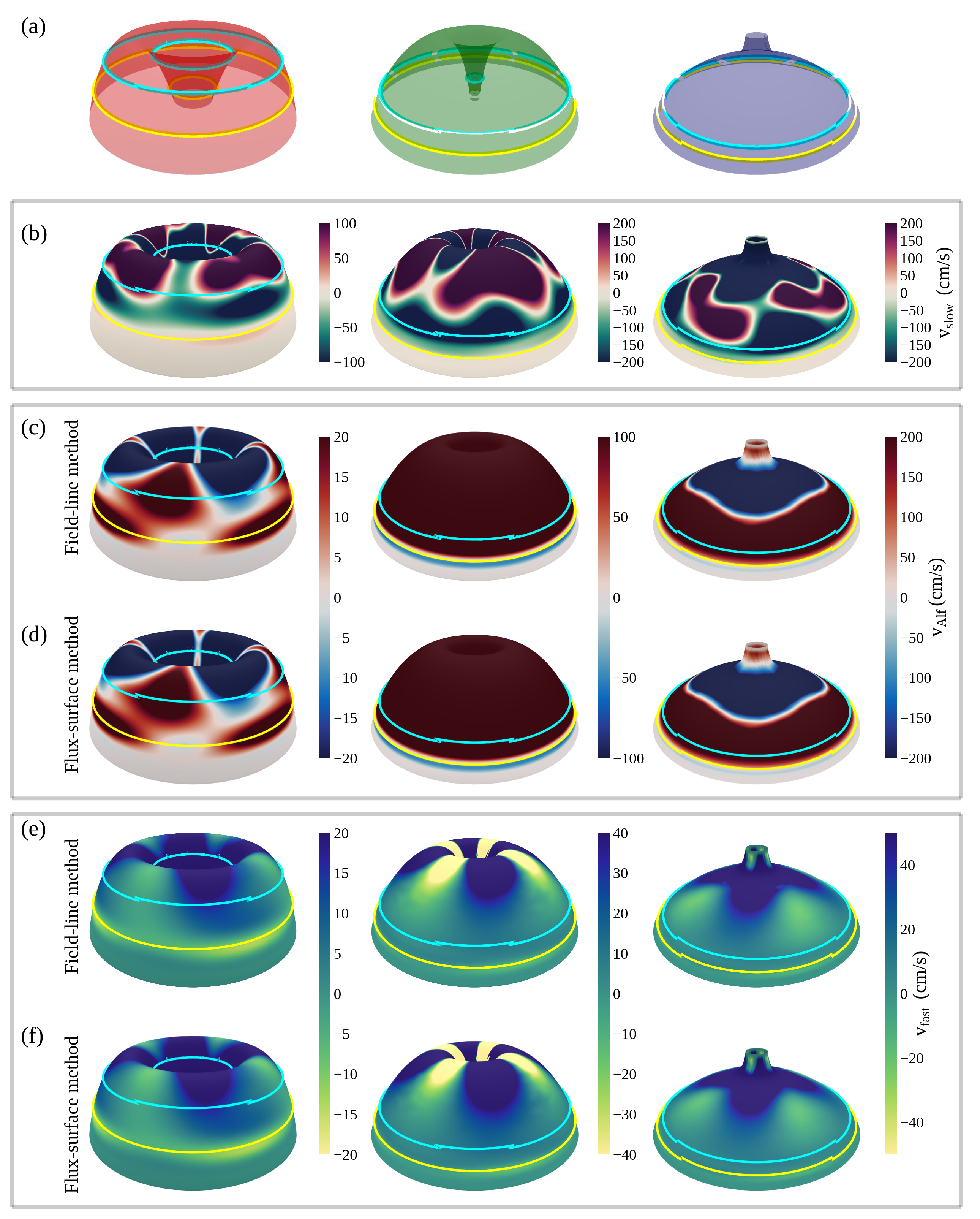}
         \caption{Comparison and validation of the proposed wave mode decomposition method with the previously used method: selected magnetic flux surfaces are shown in row (a); velocity component associated with slow magneto-acoustic wave is shown in row (b); velocity component associated with fast magneto-acoustic wave and Alfv\'en wave using our new method is shown in rows (c) and (e), respectively; velocity component associated with fast magneto-acoustic wave and Alfv\'en wave using the method used by \citet{2015ApJ...799....6M} is displayed in rows (d) and (f), respectively. Over-plotted yellow and turquoise solid line curves represent the transition region and Alfv\'en-acoustic equipartition layer, respectively.}
            \label{vel_comparison}
\end{figure*}

All of the methods discussed above  work well in special cases when a clear flux tube geometry is present or when we have prior knowledge of the nature of the wave propagation vector.
However, in general cases e.g. in realistic simulations or in real solar observations, the methods discussed above face limitations.
A method to decompose waves that can be used in more realistic scenarios is based on the compressive versus rotational nature of the velocity vectors (\citealt{2017MNRAS.466..413C,2018A&A...618A..87K,2019ApJ...883..179K}). 
In this method, using the (assumed linear) velocity from the simulations, the following quantities are calculated: 
\begin{align}
    f_{alf} &= \hat{\mathbf{s}} \cdot (\nabla \times \textbf{v}) \,, \\
    f_{slow} &= \hat{\mathbf{s}} \cdot \nabla(\textbf{v} \cdot \hat{\mathbf{s}}) \,, \\
    f_{fast} &=  \nabla \cdot (\textbf{v}-(\textbf{v} \cdot \hat{\mathbf{s}})\hat{\mathbf{s}}) \,.
\end{align}
Here, the quantities computed are not velocities but relate to their derivatives. Therefore, these proxies can not directly be used to calculate the wave energy fluxes.

\subsection{A field-line based mode decomposition}
To overcome the limitations of the methods discussed above, we suggest a new technique that is inspired by well-known Frenet–Serret formulae.
In the proposed approach, individual magnetic field-lines are used for wave identification rather than constructing constant magnetic flux surfaces. 

In this approach, just like in other decomposition strategies discussed above, the unit vector in the direction of the background magnetic field vector corresponds to the slow magneto-acoustic waves ($\hat{\mathbf{s}}$).
Then, we calculate the unit vector corresponding to the Alfv\'en wave and utilize the mutual orthogonality for computing the third unit vector for the fast magneto-acoustic wave.
To compute a uniquely defined unit vector corresponding to the Alfv\'en wave, we first construct a local plane at any point on the curved magnetic field-line by connecting it to two neighbouring points on each side of the field-line and then we calculate the unit vector normal to this plane and associate it to the Alfv\'en wave ($\hat{\bm{\phi}}$).
Since the solar atmosphere is gravitationally stratified, the magnetic field expands with height and hence magnetic field-lines always possess substantial curvature.
Finally, the unit vector corresponding to the fast magneto-acoustic wave is calculated by $\hat{\mathbf{n}}=\hat{\bm{\phi}} \times \hat{\mathbf{s}}$.
To be consistent throughout the domain, we implement a check such that our unit vector for Alfv\'en wave does not flip direction after crossing the point of inflection.
Moreover, since we wish to compare our results with the method used by \citet{2015ApJ...799....6M}, we additional require that the unit vector for fast magneto-acoustic wave is always orientated outwards of the flux surface.
A cartoon representation of this proposed method is illustrated in Fig. \ref{cartoon2}.
The advantage of this method is that whenever we have field-lines available (either numerically or analytically), it is always possible to define this frame on every point of every field-line, and we never need to resort to flux surfaces or tubes. Moreover, the projections are well defined for all curves, and can directly lead to quantification of wave energy fluxes along field-lines.

\section{Results and Discussion}\label{results}
\subsection{Equivalence of flux surface and field-line based decompositions}
To compare and validate the suggested approach of wave decomposition, we select three flux surfaces (red, green, and blue in Fig. \ref{magnetic_surfaces}) and then isolate the velocity perturbations corresponding to the three MHD waves using both the proposed `field-line based method' and the `flux-surface method' used by \citet{2015ApJ...799....6M} for a fixed time instance.
The choice of flux surfaces is based on their spatial topology with respect to the magnetic null point.
The flux surface shown in red stays quite far from the spine and fan surface of the null configuration while the surface in green reaches close to the inner spine and the fan surface.
The flux surface in blue samples parts of the outer spine and fan surface, and is closest to the magnetic null among the three selected surfaces.
The three selected magnetic flux surfaces are shown in row (a) of Fig.~\ref{vel_comparison}: note that these three flux surfaces are shown on the same scale, but their actual physical extent and relative configurations are different as seen in Fig.~\ref{magnetic_surfaces}.

In our proposed field-line based wave decomposition approach, we first need to trace the magnetic field-lines in our 3D simulation domain. 
In fact, we selected three circular seed areas at the bottom of our simulation domain ($z=0$) with varying radii and placed 500 seed points on each of them, to trace multiple field-lines.
Due to the axisymmetric nature of the background magnetic field, tracing the associated magnetic field-lines through the simulation domain provides three constant magnetic flux surfaces as shown in row (a) of Fig.~\ref{vel_comparison}. 
In general, seed points can be randomly placed however, here they are placed on a circular region to provide constant magnetic flux surfaces, allowing us to compare directly to earlier approaches that relied on calculating normal vectors of flux surfaces.
To calculate the velocity perturbation associated with slow magneto-acoustic waves, we calculate the corresponding tangent unit direction vectors. 
The component of local velocity vector in this direction constitutes the velocity perturbation associated with slow magneto-acoustic waves ($v_{slow}$).
It is shown in row (b) of Fig. \ref{vel_comparison}.
\begin{figure*}
     \centering
        \includegraphics[scale=0.44,trim=0cm 0cm 0cm 0cm]{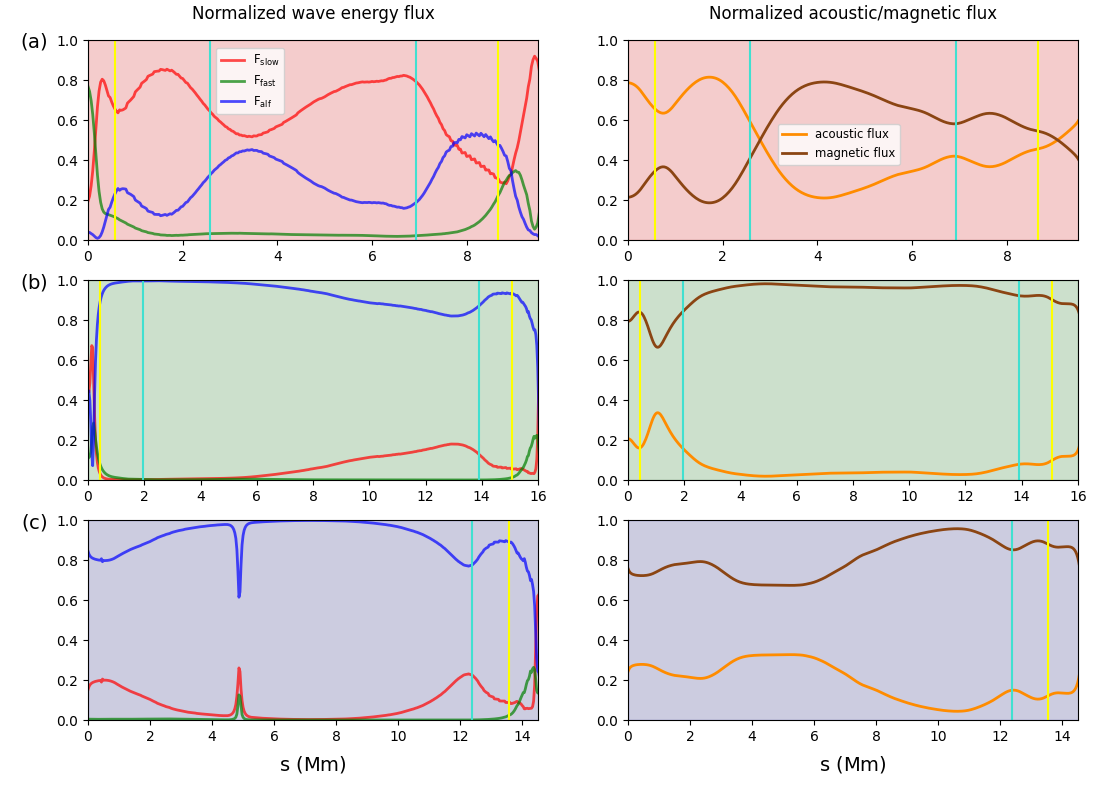}
         \caption{
         Energy flux comparison for the three selected surfaces also indicated as background colors:
         red (first row), green (second row), blue (third row). Left column: energy fluxes calculated using Eqs. (\ref{available_energy_flux1}) - (\ref{available_energy_flux3}): slow
         (red line), fast (green line) and Alfv\'en (blue line). Right column: acoustic (orange line) and magnetic (brown line) fluxes, calculated as the first and second terms in Eq. (\ref{total_wave_energy_flux_vector}), respectively.}
            \label{energy_comparison}
\end{figure*}

To isolate the other two modes in our field-line based method, we use the fact that slow and fast magneto-acoustic waves have perturbations in the plane formed by joining three consecutive points on the field-line while the Alfv\'en wave has velocity perturbations in the direction normal to this plane in which the background field is lying. 
Once the unit direction vectors have been computed, we compute the velocity perturbation associated with various waves by taking projections of velocity vector onto these directions.
Our simulations run for $\sim$ 48 minutes in total and reach a periodically repeating state after around 15 minutes, where the system is time-varying, with a quasi-steady background state and oscillations around its mean value.
In this quasi-steady state, energy input to the domain through the applied wave driver is balanced by the energy dissipation through damping at the upper absorbing layer (mimicking an upper open boundary, to some extent) and/or through numerical dissipation.
To compare and validate our proposed approach, we choose data at a selected time ($\sim$ 36 minutes).
To obtain the velocity perturbation associated with the Alfv\'en wave ($v_{alf}$) and fast magneto-acoustic wave ($v_{fast}$), velocity components along the azimuthal unit vector ($\hat{\bm{\phi}}$) and normal vector ($\hat{\mathbf{n}}$) are calculated and displayed in row (c) and row (e) of Fig. \ref{vel_comparison}, respectively.

For comparison and validation of our proposed field-line based wave identification method, we also calculated the velocity components for three MHD waves using the flux-surface method.
In this method, the direction vector for slow magneto-acoustic waves is calculated in the same way i.e. along the local magnetic field vector and is hence identical to the one shown in panels (b).
To calculate the direction vector corresponding to fast magneto-acoustic and Alfv\'en waves that lie in the plane normal to the background magnetic field vector, a small plane is constructed at each point on the magnetic flux surface, by joining it to the two adjacent points of the neighbouring field-line.
Then, the normal vector to this plane is calculated for the fast magneto-acoustic wave.
For consistency, this normal vector is always oriented away from the central axis of the background magnetic field. 
The unit vector corresponding to the Alfv\'en wave is then computed using the orthogonality condition i.e. $\hat{\bm{\phi}}=\hat{\mathbf{s}} \times \hat{\mathbf{n}}$.
Velocity perturbations along these directions are computed and shown in rows (d) and (f) of Fig.~\ref{vel_comparison} that represent velocity perturbations corresponding to Alfv\'en wave and fast magneto-acoustic waves, respectively.
Note that the selected neighbouring magnetic field-line must be part of the constant magnetic flux surface, which was easy to construct here due to axisymmetry.

If we compare panels in rows (c) and (e) that are calculated by field-line based method, with rows (d) and (f) that are calculated by flux-surface method, we see that both methods produce similar results.
Small differences between them are purely numerical and their accuracy relies on different factors, e.g. on the number of seed points chosen to construct the surface in the flux-surface method, or in our proposed method on the step-size used for spatial integration while tracing a field-line.
Thus we suggest that both our proposed method and the flux-surface method works equally well in decomposing linear MHD waves in the solar atmosphere, whenever we have a situation characterized by axisymmetric magnetic fields.
One must note that the wave decomposition used by \citet{2016ApJ...819L..11S} would also work equally well here as the magnetic field is symmetric about the $z$ axis. 
However, it would give inconsistent results if the magnetic field would possess finite twist, instead of expanding only in the vertical direction.
In axisymmetric twisted flux tube scenarios, the flux-surface method and the one proposed here would work.
However, in realistic magneto-convection simulations and in actual solar observations, magnetic fields are almost never fully axisymmetric as magnetic flux elements are continuously jostled by the convective flows.
In these circumstances, it is impossible to create a unique flux surface at any given location, however a magnetic field-line can be traced in the domain in a unique way.
Therefore, the proposed field-line based method has merits over other methods to investigate wave propagation and wave mode transformations taking place in more realistic models and in actual solar observations as it only relies on the individual magnetic field-line geometry to identify the three MHD waves.

\subsection{Analysis of wave fluxes and effect of the null point}
To quantify the energy content in the various MHD modes, we take a spatial average of the energy fluxes over all the field-lines constituting the three selected surfaces and then additionally average over three wave periods.
In ideal linear MHD, the wave energy density is a conserved quantity and is related with the wave energy flux of the following form (\citealt{1974soch.book.....B, 1985GApFD..32..123L, 2003ApJ...599..626B}):
\begin{align}
    \textbf{F}_{\mathrm{waves}}  &= p_1 \textbf{v} + \textbf{B}_1 \times (\textbf{v} \times \textbf{B}_0)  \,,
    \label{total_wave_energy_flux_vector}
\end{align}
where the quantities with subscript '1' represent the perturbation part that evolve during the simulation while the quantities with subscript '0' correspond to the steady background. 
The first term on the right hand side of the above equation represents the acoustic flux, while the second term represents the Poynting or magnetic flux. 

In a low beta plasma, found in the majority of our simulation domain and in the (upper) solar chromosphere and corona throughout, the slow magneto-acoustic waves are predominantly of acoustic nature while fast magneto-acoustic \textbf{waves} and Alfv\'en waves are of magnetic nature. 
Therefore, comparing the acoustic and magnetic wave energy fluxes separates the slow wave energy flux from the combined energy flux associated with fast and Alfv\'en waves. However, this does not provide any insight about the energy flux distribution among fast magneto-acoustic \textbf{waves} and Alfv\'en waves.
Moreover, it is important to separate out the energy flux contribution from different waves in order to estimate the available energy fluxes in these waves and to investigate energy transfer among various wave modes in the process of mode transformation.
For example, in case of standing waves, the total acoustic/magnetic flux might be zero but individual wave energy fluxes will always give positive quantities with no cancellation (\citealt{2012ApJ...758...96F}).

Therefore, to get an estimate of the contribution of individual MHD modes, we calculate the energy flux contained in the three different MHD waves using the following relations: 
\begin{align}
    \label{available_energy_flux1}
    F_{\mathrm{slow}} &= \rho v^2_{slow} c_s \,,\\
    F_{\mathrm{Alf}} &= \rho v^2_{alf} v_A \,,\\
    F_{\mathrm{fast}} &= \rho v_{fast}^2 \sqrt{c_s^2+v_A^2}\,.
    \label{available_energy_flux3}
\end{align}
These expressions just combine the ram pressure in the decomposed wave modes (identified from the MHD mode decomposition) with the relevant group speed of the mode in question: in a low-beta plasma, this group speed is the sound speed for the slow, the Alfv\'en speed for the Alfv\'en mode, and the combined sound-Alfv\'en speed for the fast mode. 

\begin{figure}
     \centering
        \includegraphics[scale=0.29]{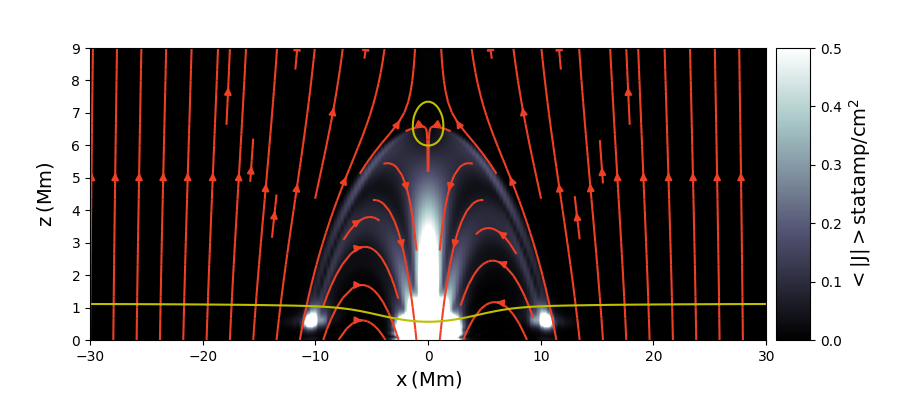}
         \caption{Spatial map of magnitude of current (time averaged over three wave periods) shown for a $y=0$ slice through the domain.}
            \label{current}
\end{figure}

In a few recent works, the total wave energy flux vector from Eq.~(\ref{total_wave_energy_flux_vector}) is decomposed into three components by taking projections in the same way as done for velocity components associated with the different MHD wave modes (\citealt{2015ApJ...799....6M,2018MNRAS.480.2839L}).
However, the wave energy flux component computed in this way is not always a correct representation and should be interpreted with caution.
For slow magneto-acoustic waves, wave energy can be computed as the projection of the acoustic component of the total wave energy flux vector along the magnetic field vector, as it travels predominantly along the magnetic field in a low beta plasma.
However, the parallel component of the total wave energy flux vector also gets a contribution from the magnetic component associated with Alfv\'en and fast magneto-acoustic waves and therefore does not necessarily quantify the purely slow wave energy flux.
In order to be able to separate the net wave energy flux associated with the different MHD waves, knowledge of the wave propagation vector is a pre-requisite, which is not known because fast magneto-acoustic waves travel (nearly) isotropically originating from various possible sources. 
Therefore, in the present work, we restrict ourselves to calculating the ``available'' energy flux using the expressions given in Eqns.~(\ref{available_energy_flux1} - \ref{available_energy_flux3}).

The three rows in Fig.~\ref{energy_comparison} shows the energy flux comparison for the three selected surfaces as can be distinguished by the background color. 
The panels in the left column compare the ``available'' wave energy flux corresponding to the different MHD waves (normalized to the sum of the wave energy fluxes at that location) while the panels at right compare magnitudes of acoustic and magnetic parts of the wave energy flux vector $\mathbf{F}_{\mathrm{wave}}$ (normalized to the sum of acoustic and magnetic flux at that location). 
The color coding for various curves is the same for the three panels in each column (according to the inserted legends in row (a)). 
Another point to notice is that the length of the horizontal axis corresponds to the total length of the field-line constituting the selected surface, which hence varies from row to row. 
The right end of the horizontal axis always corresponds to the end of the field-line that is farthest from the axis while lying in the $z=0$ plane.
The left end of the axis i.e. $\mathrm{s}=0$ corresponds to $z=0$ for red and green surface, while for blue surface it corresponds to $z=10$ Mm i.e. top of the simulation domain.
The vertical lines represent the locations of the Alfv\'en-acoustic equipartition layer (yellow) and the transition region (turquoise).
There are two vertical yellow and two vertical turquoise lines for red and green panels as the field-lines return back to the surface and encounter the transition region and Alfv\'en-acoustic equipartition layer twice.
While for the blue surface, there is only one vertical yellow and one vertical turquoise line, as the field-lines for this surface do not return back to the surface and end on the top of the simulation box.
Here, we must note that although the blue surface crosses very close to the null, it does not cross the equipartition layer surrounding the null.
Since the wave decomposition method proposed here is not fully applicable for the high beta regions, we restrict our discussion of wave energy fluxes to the regions confined within the yellow vertical lines for red and green surfaces, while for the blue surface, we consider the region on the left of the yellow line.

Row (a) shows that the energy flux is distributed among magneto-acoustic waves and Alfv\'en waves as the applied driver is not necessarily in the direction normal to the magnetic field and contains both magnetic and acoustic flux. 
Since the fast wave energy flux is expected to get focused at the null point, there is not much fast wave energy flux on this surface as it is far from the null point (more discussion on selective fast wave energy accumulation follows). 

Row (b) displays the results for the green surface which reaches closer to the null point and the fan surface than the red surface.
Here in the left panel, we see that the energy flux is dominated by the Alfv\'en wave and consistent with that, the right panel is dominated by magnetic flux.
Accumulation of Alfv\'en waves at the fan surface has been previously shown by both analytical and numerical investigations where twist propagates towards a null point along the spine axis (\citealt{2003JGRA..108.1042G, 2013A&A...558A.127T}). 
These studies, on the other hand, were constrained to a uniform background density, and they solved the ideal MHD equations with the $\beta = 0$ approximation.
Similar results are found for the blue surface shown in row (c) of Fig.~\ref{energy_comparison}.
Just like the green surface, a large part of this surface is close to the fan surface, and hence is dominated by Alfv\'en wave energy flux in the left panel and by magnetic flux in the right panel.
Interestingly, we see a sharp increase in the slow and fast wave energy fluxes on the left panel at around s = 5 Mm, and an increase in the acoustic flux in the right panel at the location of closest proximity to the null point. 

Since the Alfv\'en speed drops down to nearly zero near the null point, Alfv\'en waves cease to propagate and spread over the fan surface as magnetic field-lines diverge. This explains the sharp dip in Alfv\'en energy flux when the blue surface is in close proximity for the null point. 
Moreover, a fast wave undergoes refraction due to an inhomogeneous Alfv\'en speed and gets focused near the null point. 
This focusing results in an increase of fast wave energy flux at the null point clearly seen here for the blue surface.
This fast wave accumulation near the null point can also be seen in the rows (e) and (f) of Fig.~\ref{vel_comparison}; especially in the panels corresponding to the blue surface as it is closest to the null point.
Increase in the slow wave energy flux can be a possible result of mode conversion taking place at the Alfv\'en-acoustic equipartition layer surrounding the null point (\citealt{2004A&A...420.1129M}).
Thus the obtained results here are consistent with previous analytical and numerical investigations, showing the wave accumulation at predictable locations of the magnetic topology (\citealt{2006A&A...459..641M,2013A&A...558A.127T}). We hence note that our quantification of the wave energy fluxes by the two means shown in Fig.~\ref{energy_comparison} provide complimentary info on the true wave transformations occurring throughout the domain.

\subsection{Current density accumulation}
Though we did not include resistivity in our model, to gain a notion of the locations where preferential heating might take place in such a system, we calculate current density throughout the domain.
Fig. \ref{current} displays the magnitude of the current density vector for a slice ($y=0$) that is calculated by taking a time-average over three wave periods.
We find strong current buildup at the spine and also at the fan surface. 
This is consistent with the wave energy analysis, as Alfv\'en waves spread along the fan surface, and create gradients of magnetic field that result in strong currents.
Similar results were obtained by \citet{2013A&A...558A.127T} in which they investigated behavior of Alfv\'en waves near a null point in a comparatively simpler model by ignoring gravitational stratification and acoustic effects.
This increase in current is extremely important, because it suggests that finite resistivity will eventually dissipate the wave energy and contribute to plasma heating at these locations in the solar corona.

Many of the previous models showed preferential current accumulation at the null point (\citealt{2004A&A...420.1129M,2006A&A...459..641M}) that is not present in our simulations.
The reason for this lack of current accumulation at the null point in our case is related to the nature of our applied wave driver. 
Our driver is predominately of Alfv\'enic nature and has very little wave power in the fast mode, as can be seen in the left columns of Fig. \ref{energy_comparison}.
In the earlier studies, their wave driver is in the form of fast waves, and wave energy gets focused at the null point and leads to current accumulation close to the null. 
These findings imply that in the actual solar case, where waves are stochastically excited due to turbulent convection, the nature of the excited waves cannot be predicted in advance, and thus current accumulation and plasma heating can occur in different characteristic regions of a 3D null topology, such as the null, spine, or fan plane.

\section{Conclusions}\label{conclusion}
We performed three-dimensional ideal MHD simulations of waves travelling through a gravitationally stratified solar atmosphere extending from the solar surface to the solar corona.
The magnetic field configuration is chosen to be representative of a uni-polar magnetic field region with a parasitic magnetic patch of opposite polarity at its center. This creates a null point in the corona, and such magnetic configurations are believed to be common in the lower solar atmosphere.
To study the propagation of waves excited by ubiquitous vortex flows found at the solar surface, we employ a rotational velocity driver at the bottom boundary of our simulation domain, and place it in the center of the parasitic polarity region.
Most of our simulation domain contains low-beta plasma where the slow magneto-acoustic wave is predominantly acoustic and the fast magneto-acoustic wave is predominantly magnetic in nature. However, we do have the sound-Alfv\'en equipartition height in our chromosphere, and the coronal null introduces another equipartition surface centered on this null. We introduced and applied an MHD wave decomposition method to see how wave energy fluxes traverse this fairly realistic stratified magnetic topology. 

In a fully three-dimensional setup, the magnetic component of the total wave energy flux vector also gets a contribution from Alfv\'en waves and hence, it is not always easy to separate out the wave energy fluxes for all three different MHD waves.
To tackle this issue, various wave identification methods have been used in the literature but they have certain limitations with respect to the magnetic field configuration and/or the wave propagation vector.
Here we present another wave identification method that can avoid these limitations, so that different wave modes can be isolated in more general scenarios.
Our proposed method makes use of the local Frenet frame vectors on a magnetic field-line to calculate the components of velocity perturbations corresponding to the three different MHD waves.
We selected an axially symmetric magnetic field configuration so that we can compare our proposed method with an existing method that requires to construct or have knowledge of constant and nested magnetic flux surfaces.

For comparing our method and the flux-surface method, we selected three constant flux surfaces in our domain and compared the velocity components corresponding to the different MHD modes.
We showed that both methods give identical results, as expected in this setting. However, the proposed method has advantages over the flux-surface method as it can be applied to more realistic scenarios where the magnetic field configuration does not have an axis of symmetry.
We must note that the proposed field-line based method is only applicable for identifying pure linear MHD waves in low-beta plasma regions.
However, in the proximity of the null point, magnetic gradients are steep leading to strong variations in the phase speeds of the waves, thus modifying the propagation of MHD waves. Moreover, plasma beta is more than unity near the null point and thus the proposed wave decomposition does not hold true.
We further calculated the time-averaged (over three full time period of the input driver) wave energy fluxes associated with slow, fast and Alfv\'en waves after the simulations reached a stationary state. 
We found that fast wave energy flux and Alfv\'en wave energy flux accumulate at predictable locations in the null configuration, i.e. near the null for fast waves and near the fan surface for Alfv\'en waves.
Thus, we show that even in a more realistic setup, the obtained results are consistent with previous analytical and simulation studies that used several approximations in terms of background parameters and/or ignored acoustic effects.
We found current buildup at the spine and fan surface of a null topology.
In the presence of finite resistivity, excessive currents can lead to preferential heating at specific locations and thus may have important implications in triggering solar jets (\citealt{2021RSPSA.47700217S}).

In the present simulations, we have a realistic solar atmosphere in the sense that excited waves pass through various mode transformation and reflecting layers, viz. the acoustic cutoff layer, two Alfv\'en-acoustic equipartition layers (one near the surface and one surrounding the null) and also the transition region.
Therefore, several interesting mode conversion, reflection and transmission processes take place at these layers.
The amplitude of our wave driver was intentionally kept small in order to investigate the wave propagation and transformations in a linear regime.
We have compared the available energy fluxes in various MHD waves here, however, we will further investigate the mode transformations in detail in a follow-up study and compare them with a nonlinear case. 
Also, in the nonlinear case for the present setup additional wave excitation sources might play an important role due to nonlinear effects.
We also plan to extend the presented study by including different profiles of wave drivers in terms of their spatial localization, frequency and polarization.
Since the proposed method of wave decomposition can be applied to magnetic fields obtained by extrapolations of photospheric magnetograms, we plan to perform wave studies using a variety of wave drivers on extrapolated magnetic fields that host a magnetic null.
Moreover, in the present simulations we excited waves with a fixed frequency, while it is well-known that waves excited by turbulent convection and granular buffeting cover a broadband spectrum. 
Therefore, including multi-frequency or stochastic drivers in the present setup can also be an interesting follow-up study. 
We also intend to apply the proposed wave decomposition technique in realistic magnetoconvection simulations. As shown by \citet{2021A&A...645A...3Y} slow magneto-acoustic waves carry enough wave energy flux to heat the solar chromosphere. 
However, these waves get dissipated in the chromosphere due to shock formation and do not contribute much to the higher layers. 
Whereas, magnetic waves propagate to higher layers and may heat the solar corona. Employing the wave identification method proposed here, we intend to calculate energy flux associated with fast magneto-acoustic and Alf\'en waves in such realistic simulations and estimate their energy content.

\begin{acknowledgements}
NY and RK are supported by Internal funds KU Leuven, project C14/19/089 TRACESpace. RK further received funding from the European Research Council (ERC) under the European Union’s Horizon 2020 research and innovation programme (grant agreement no. 833251 PROMINENT ERC-ADG 2018),a joint FWO-NSFC grant G0E9619N and FWO grant G0B4521N. BP is supported by the FWO grant 1232122N.
\end{acknowledgements}
 \bibliographystyle{aa} 
 \bibliography{null} 

\end{document}